\documentclass[prd,superscriptaddress,twocolumn,nofootinbib,showpacs,preprintnumbers,floatfix]{revtex4}

\usepackage{mathrsfs}
\usepackage{amsfonts,amssymb,amsmath}
\usepackage{graphicx,epsfig}
\usepackage{multirow}
\usepackage{verbatim}

\def\fsu5{$\cal{F}$-$SU(5)$}

\def\m1half{$M_{1/2}$}
\def\m3half{$M_{3/2}$}
\def\m32{$M_{32}$}

\begin{document}

\title{A Higgs Mass Shift to 125~GeV and A Multi-Jet Supersymmetry Signal: \\ Miracle of the Flippons at the $\mathbf{\sqrt{s} = 7}$~TeV LHC}

\author{Tianjun Li}

\affiliation{State Key Laboratory of Theoretical Physics, Institute of Theoretical Physics,
Chinese Academy of Sciences, Beijing 100190, P. R. China }

\affiliation{George P. and Cynthia W. Mitchell Institute for Fundamental Physics and Astronomy,
Texas A$\&$M University, College Station, TX 77843, USA }

\author{James A. Maxin}

\affiliation{George P. and Cynthia W. Mitchell Institute for Fundamental Physics and Astronomy,
Texas A$\&$M University, College Station, TX 77843, USA }

\author{Dimitri V. Nanopoulos}

\affiliation{George P. and Cynthia W. Mitchell Institute for Fundamental Physics and Astronomy,
Texas A$\&$M University, College Station, TX 77843, USA }

\affiliation{Astroparticle Physics Group, Houston Advanced Research Center (HARC),
Mitchell Campus, Woodlands, TX 77381, USA}

\affiliation{Academy of Athens, Division of Natural Sciences,
28 Panepistimiou Avenue, Athens 10679, Greece }

\author{Joel W. Walker}

\affiliation{Department of Physics, Sam Houston State University,
Huntsville, TX 77341, USA }


\begin{abstract}
We describe a model named No-Scale \fsu5 which is simultaneously capable of explaining the dual signals emerging
at the LHC of i) a 124--126 GeV Higgs boson mass $m_h$, and ii) tantalizing low-statistics excesses in the multi-jet data
which may attributable to supersymmetry.  These targets tend to be mutually exclusive in more conventional approaches.
The unified mechanism responsible for both effects is the introduction of a rather unique set of
vector-like multiplets at the TeV scale, dubbed {\it flippons}, which i) can elevate $m_h$ by around 3-4 GeV
via radiative loop corrections, and ii) flatten the running of the strong coupling and color-charged gaugino, resulting
in a prominent collider signal from production of light gluino pairs.  This well motivated theoretical framework maintains consistency
with all key phenomenological constraints, and all residual parameterization freedom may in principle be fixed by a
combination of the two experiments described.
We project that the already collected luminosity of $5~{\rm fb}^{-1}$ may be sufficient to definitively
establish the status of this model, given appropriate data selection cuts.
\end{abstract}


\preprint{ACT-22-11, MIFPA-11-55}

\maketitle


\section{Introduction}

\subsection{The Two-Pronged LHC Mission}

The chiral electroweak (EW) $SU(2)_{\rm L} \times U(1)_{\rm Y}$ gauge symmetry of the standard model (SM)
prohibits the expression of tree-level fermionic Dirac masses, while the massless chiral symmetry
limit simultaneously softens the na\"ively expected linear divergence of fermionic loops to
be instead manifest in logarithms.  Exact gauge symmetry protects the masses of non-Abelian
force-carriers to all orders.  Having nullified the zeroth order masses,
Planck-small mass terms may be transferred out of symmetry-preserving interactions with a
fundamental scalar ``Higgs'' field during a radiatively triggered spontaneous breakdown of the
$SU(2)_{\rm L}\times U(1)_{\rm Y}$ vacuum.  By coupling every SM field to a half-integrally spin-shifted partner,
supersymmetry (SUSY) extends the favorable fermionic birth-right to fields of the bosonic sector,
generating precise counter-terms to problematic loops via the circulation of partners bearing opposite
spin-statistics, and taming the catastrophic quadratic divergence of the scalar Higgs field
itself.  A dual search for these ostensibly independent mechanisms constitutes
the nominal {\it raison d'\^etre} of the Large Hadron Collider (LHC).

\subsection{The LHC Higgs Search Strategy}

The LHC strategy for the hunt of the SM Higgs boson:
i) is based upon the gluon pair fusion to Higgs triangle diagram calculated in 1978
by Georgi $\textit{et al.}$ (GGMN)~\cite{Georgi:1977gs},
which is the leading production mechanism by a merit factor of $\cal{O}$(10),
and ii) prominently features the Higgs to two gamma triangle diagram calculated in 1976 by
Ellis $\textit{et al.}$ (EGN)~\cite{Ellis:1975ap}, which is
the cleanest decay mode in the relevant mass range.
In the minimal supersymmetric SM extension (MSSM), holomorphy of the
superpotential and anomaly cancellation for the fermionic
Higgsinos each imply that independent complex $SU(2)_{\rm L}$ Higgs doublets
must separately provide mass to the up-like and down-like fields.  The
resulting eight real degrees of freedom are reduced by three in the transfer
of a longitudinal polarization to the massive $W^\pm$ and $Z^0$ bosons,
leaving five massive physical Higgs fields after symmetry breaking.
Moreover, non-SM fields will circulate around both the GGMN production triangle and
the EGN decay triangle, in principle modifying both the Higgs production cross section
and decay width and mode, and thus also impacting the observed detection limits.
Fortunately, the extra contributions are generally negligible at the leading 
precision.  It is generally expected that the lightest CP-even field $h$
would manifest itself similarly to the single physical Higgs of the SM,
although there remain notable distinctions, including an explicit formula
for the tree level mass-splitting, which indicates that the light Higgs
must have a mass $m_h$ below that of the $Z$-boson at around $91$~GeV.
Such a light field would have easily been detected at LEP, which has placed
a lower bound on the MSSM Higgs of $114$~GeV.  However, a critical
loophole allows SUSY, and even the minimal MSSM formulation, to escape
the null result intact; radiative contributions to the squared Higgs mass
$m_h^2$ from the extremely heavy (and thus strongly Higgs-coupled) top
quark are sufficient to lift the SUSY construction out of danger.

\subsection{The Miracle of the Flippons}

In reality, specific model-based predictions for the Higgs and SUSY structure
are intimately correlated, as we shall demonstrate within the context of a particular phenomenologically favorable
construction named No-Scale \fsu5~\cite{Li:2010ws, Li:2010mi,Li:2010uu,Li:2011dw, Li:2011hr, Maxin:2011hy,
Li:2011xu, Li:2011in,Li:2011gh,Li:2011rp,Li:2011fu,Li:2011ex,Li:2011av}, which is defined by the convergence of the
${\cal F}$-lipped $SU(5)$~\cite{Barr:1981qv,Derendinger:1983aj,Antoniadis:1987dx} grand unified theory (GUT),
two pairs of hypothetical TeV scale vector-like supersymmetric multiplets, dubbed {\it flippons}, with origins in
${\cal F}$-theory~\cite{Jiang:2006hf,Jiang:2009zza,Jiang:2009za,Li:2010dp,Li:2010rz}, and the dynamically
established boundary conditions of No-Scale Supergravity~\cite{Cremmer:1983bf,Ellis:1983sf, Ellis:1983ei, Ellis:1984bm, Lahanas:1986uc}.
Careful numerical analysis of the viable No-Scale \fsu5 parameter space
yields a prediction for $m_h$ in the range of 119.0~GeV to 123.5~GeV~\cite{Li:2011xg},
consistent with limits from the CMS~\cite{PAS-HIG-11-022}, ATLAS~\cite{ATLAS-CONF-135,ATLAS:2011ww},
CDF and D\O~Collaborations~\cite{:2011ra}.  However, in light of the most
recent CMS and ATLAS reports~\cite{Collaboration:2012tx,Collaboration:2012si} of a Higgs
signal at 124--126 GeV with independent local significances greater than $3 \sigma$ over
background (or around $2 \sigma$ each after global ``look elsewhere'' compensation),
there is a critical modification that we now entertain.  By
coupling to the Higgs, the vector-like {\it flippon} multiplet will itself
have an impact on the predicted value of $m_h$~\cite{Moroi:1992zk,Babu:2008ge}.  Just as radiative top-quark loops once
elevated $m_h$ to save supersymmetry (and were in fact taken as suggestive evidence for a heavy
top quark prior to the observation by D\O ~and CDF), radiative loops in heavy vector-like
multiplets may provide a second such boost, but this time to save No-Scale \fsu5,
and to provide suggestive evidence for the new {\it flippon} field.

We will demonstrate explicitly that No-Scale \fsu5 may convincingly explain 
this exciting new signal, while simultaneously accounting for tantalizing
excesses at lower statistical significance in the search for SUSY via
multi-jet production at ATLAS~\cite{Aad:2011qa} and CMS~\cite{PAS-SUS-09-001}.
This is particularly noteworthy, because the two results will tend to be
mutually exclusive for more conventional MSSM constructions.  In particular,
the mechanism for elevation of the Higgs mass will typically correspond to
squark and gluino masses which are far too heavy to have yet peeked above
the SM background for the initial $\sqrt{s} = 7$~TeV operating phase of the LHC.
No-Scale \fsu5 takes advantage of the same strongness of the Higgs-top quark coupling
which provides the primary lifting of the SUSY Higgs mass to generate a hierarchically light
partner stop in the SUSY mass-splitting.  However, this rather generic mechanism is not
in itself enough.  The model further leverages the same vector-like multiplets which provide
the secondary Higgs mass perturbation to flatten the renormalization group equation (RGE)
running of universal color-charged gaugino mass, blocking the standard logarithmic
enhancement of the gluino mass at low energies, and producing the distinctive mass
ordering $M({\widetilde{t_1}}) < M({\widetilde{g}}) < M({\widetilde{q}})$ of a light stop
and gluino, both substantially lighter than all other squarks. 
We will demonstrate that the consequence of this spectrum is a spectacular signal of
SUSY events in the ultra-high jet multiplicity channels, which is not just in passive compliance
with LHC production limits, but is moreover an active enhancement of the theoretical and experimental accord. 
We project that the already collected luminosity of $5~{\rm fb}^{-1}$ may be sufficient to definitively
establish the status of this SUSY signature, given appropriate data selection cuts.
At this intersection of the two great LHC causes, we moreover find the parameterization
freedom of the model to be exhausted in a manner that is profoundly consistent with all
existing phenomenological constraints.  This is the {\it Miracle of the Flippons}.


\section{The No-Scale $\mathbf{\cal{F}}$-$\mathbf{SU(5)}$ Model}

The No-Scale \fsu5 construction~\cite{Li:2010ws, Li:2010mi,Li:2010uu,Li:2011dw, Li:2011hr, Maxin:2011hy,
Li:2011xu, Li:2011in,Li:2011gh,Li:2011rp,Li:2011fu,Li:2011ex,Li:2011av} derived from local F-Theory model building inherits all of the most beneficial
phenomenology of flipped $SU(5)$~\cite{Nanopoulos:2002qk,Barr:1981qv,Derendinger:1983aj,Antoniadis:1987dx},
including fundamental GUT scale Higgs representations (not adjoints), natural doublet-triplet
splitting, suppression of dimension-five proton decay
and a two-step see-saw mechanism for neutrino masses,
as well as all of the most beneficial theoretical motivation of No-Scale
Supergravity~\cite{Cremmer:1983bf,Ellis:1983sf, Ellis:1983ei, Ellis:1984bm, Lahanas:1986uc},
including a deep connection to string theory in the infrared limit,
the natural incorporation of general coordinate invariance (general relativity),
a mechanism for SUSY breaking which preserves a vanishing cosmological constant at the tree level
(facilitating the observed longevity and cosmological flatness of our Universe~\cite{Cremmer:1983bf}),
natural suppression of CP violation and flavor-changing neutral currents, dynamic stabilization
of the compactified spacetime by minimization of the loop-corrected scalar potential and a dramatic
reduction in parameterization freedom.

The dimension five proton decays mediated by colored Higgsinos are highly suppressed due to the missing partner mechanism and TeV-scale $\mu$ term. However, the dimension five proton decay non-renormalizable operators suppressed by the Planck scale generically give the very fast proton decays, which is a well known problem in the supersymmetric GUTs. Interestingly, in the string model building, there exists at least one anomalous $U(1)_X$ gauge symmetry, which may be used to suppress those dimension five proton decay operators~\cite{Harnik:2004yp}.

Written in full, the gauge group of Flipped $SU(5)$ is $SU(5)\times U(1)_{X}$, which can be embedded into $SO(10)$.
The generator $U(1)_{Y'}$ is defined for fundamental five-plets as $-1/3$ for the triplet members, and $+1/2$ for the doublet.
The hypercharge is given by $Q_{Y}=( Q_{X}-Q_{Y'})/5$.  There are three families of Standard Model (SM) fermions,
whose quantum numbers under the $SU(5)\times U(1)_{X}$ gauge group are
\begin{equation}
F_i={\mathbf{(10, 1)}} \quad;\quad {\bar f}_i={\mathbf{(\bar 5, -3)}} \quad;\quad {\bar l}_i={\mathbf{(1, 5)}},
\label{eq:smfermions}
\end{equation}
where $i=1, 2, 3$.  There is a pair of ten-plet Higgs for breaking the GUT symmetry, and a pair
of five-plet Higgs for electroweak symmetry breaking (EWSB).
\begin{eqnarray}
& H={\mathbf{(10, 1)}}\quad;\quad~{\overline{H}}={\mathbf{({\overline{10}}, -1)}} & \nonumber \\
& h={\mathbf{(5, -2)}}\quad;\quad~{\overline h}={\mathbf{({\bar {5}}, 2)}} &
\label{eq:Higgs}
\end{eqnarray}
Since we do not observe mass degenerate superpartners for the known SM fields, SUSY must itself be broken around the TeV scale.
In the minimal supergravities (mSUGRA), this occurs first in a hidden sector, and the secondary propagation by gravitational interactions
into the observable sector is parameterized by universal SUSY-breaking ``soft terms'' which include the gaugino mass $M_{1/2}$, scalar mass
$M_0$ and the trilinear coupling $A$.  The ratio of the low energy Higgs vacuum expectation values (VEVs) $\tan \beta$, and the sign of
the SUSY-preserving Higgs bilinear mass term $\mu$ are also undetermined, while the magnitude of the $\mu$ term and its bilinear soft term $B_{\mu}$
are determined by the $Z$-boson mass $M_Z$ and $\tan \beta$ after EWSB.  In the simplest No-Scale scenario,
$M_0$=A=$B_{\mu}$=0 at the unification boundary, while the complete collection of low energy SUSY breaking soft-terms evolve down 
with a single non-zero parameter $M_{1/2}$.  Consequently, the particle spectrum will be proportional to $M_{1/2}$ at leading order,
rendering the bulk ``internal'' physical properties invariant under an overall rescaling.  The matching condition between the low-energy value of
$B_\mu$ that is demanded by EWSB and the high-energy $B_\mu = 0$ boundary is notoriously difficult to reconcile under the
RGE running.  The present solution relies on modifications to the $\beta$-function coefficients that are generated by the {\it flippon}
loops.  

Naturalness in view of the gauge hierarchy and $\mu$ problems suggests that the {\it flippon} mass $M_{\rm V}$ should be of the TeV order.
A ``$\mu$'' term for the {\it flippon} masses can be forbidden by either a discrete or continuous symmetry.  Also, we may only have trilinear Yukawa terms in the superpotential  in the usual string model building.  The {\it flippon} masses can be generated
via invocation of i) the Giudice-Masiero mechanism~\cite{Giudice:1988yz,Babu:2002tx}, where a suitable ``$\mu$'' term may be generated from
high-dimensional operators, or ii) additional F-theoretic Standard Model singlets, to which the {\it flippons} may couple and subsequently
obtain a mass as those singlets acquire VEVs.  The latter scenario is similar to the solution to the ``$\mu$'' problem in the next-to-the-Minimal
Supersymmetric Standard Model (NMSSM).  Avoiding a Landau pole for the strong coupling constant restricts the set of vector-like multiplets which may be
given a mass in this range to only two constructions with flipped charge assignments, which have been explicitly realized
in the $F$-theory model building context~\cite{Jiang:2006hf,Jiang:2009zza, Jiang:2009za}.  We adopt the multiplets
\begin{eqnarray}
& {XF} = {\mathbf{(10,1)}} \equiv (XQ,XD^c,XN^c) ~;~ {\overline{XF}} = {\mathbf{({\overline{10}},-1)}} & \nonumber \\
& {Xl} = {\mathbf{(1, -5)}} \quad;\quad{\overline{Xl}} = {\mathbf{(1, 5)}}\equiv XE^c \, , &
\label{eq:flippons}
\end{eqnarray}
where $XQ$, $XD^c$, $XE^c$ and $XN^c$ carry the same quantum numbers as the quark doublet, right-handed down-type quark,
charged lepton and neutrino, respectively.  Alternatively, the pair of $SU(5)$ singlets may be discarded, but phenomenological consistency then
requires the substantial application of unspecified GUT thresholds.  In either case, the (formerly negative) one-loop $\beta$-function
coefficient of the strong coupling $\alpha_3$ becomes precisely zero, flattening the RGE running, and generating a wide
gap between the large $\alpha_{32} \simeq \alpha_3(M_{\rm Z}) \simeq 0.11$ and the much smaller $\alpha_{\rm X}$ at the scale $M_{32}$ of the intermediate
flipped $SU(5)$ unification of the $SU(3)_C \times SU(2)_{\rm L}$ subgroup.  This facilitates a very significant secondary running phase
up to the final $SU(5) \times U(1)_{\rm X}$ unification scale~\cite{Li:2010dp}, which may be elevated by 2-3 orders of magnitude
into adjacency with the Planck mass, where the $B_\mu = 0$ boundary condition fits like hand to glove~\cite{Ellis:2001kg,Ellis:2010jb,Li:2010ws}.  
This natural resolution of the ``little hierarchy'' problem corresponds also to true string-scale gauge coupling unification in
the free fermionic string models~\cite{Jiang:2006hf,Lopez:1992kg} or the decoupling scenario in F-theory models~\cite{Jiang:2009zza,Jiang:2009za},
and also helps to address the monopole problem via hybrid inflation.

A majority of the bare-minimally constrained~\cite{Li:2011xu} No-Scale $\cal{F}$-$SU(5)$ parameter space depicted
in Figure~(\ref{fig:Higgs_wedge}), which is defined by simultaneous consistency with
i) the dynamically established high-scale boundary conditions $M_0$=A=$B_{\mu}$=0 of No-Scale Supergravity,
ii) radiative electroweak symmetry breaking,
iii) precision LEP constraints on the lightest CP-even Higgs boson $m_{h}$~\cite{Barate:2003sz,Yao:2006px} and other light SUSY chargino and neutralino mass content,
iv) the world average top-quark mass $172.2~{\rm GeV} \leq m_{\rm t} \leq 174.4~{\rm GeV}$, and
v) a single, neutral supersymmetric cold dark-matter (CDM) candidate providing a relic density within 
the 7-year WMAP limits $0.1088 \leq \Omega_{\rm CDM} \leq 0.1158$~\cite{Komatsu:2010fb},
remains viable even after careful comparison against the first inverse femtobarn of LHC data~\cite{Li:2011fu,Li:2011av}.
Moreover, a highly favorable ``golden'' subspace~\cite{Li:2010ws,Li:2010mi,Li:2011xg} exists which may simultaneously account for the key rare process
limits on the muon anomalous magnetic moment $(g~-~2)_\mu$ and the branching ratio of the flavor-changing neutral current decays
$b \to s\gamma$ and $B_{s}^{0} \to \mu^+\mu^-$.  The intersection of these experimental bounds is highly non-trivial,
as the tight theoretical constraints, most notably the vanishing of $B_\mu$ at the high scale boundary, render the residual
parameterization deeply insufficient for arbitrary tuning of even isolated predictions, let alone the union of all predictions.

In addition, a top-down consistency condition on the gaugino boundary mass $M_{1/2}$ is dynamically determined at a secondary local minimization of
the minimum of the Higgs potential $V_{\rm min}$, which is demonstrably consistent with the bottom-up phenomenological
approach ~\cite{Li:2010uu,Li:2011dw,Li:2011ex}.  This fixing of the supermultiplet $F$-terms for the compactification radii $R \sim 1/M_{\rm String}\propto 1/M_{1/2}$ in terms of the mass scale $M_{1/2}$ is analogous to the fixing of the Bohr atomic radius $a_0 = 1/(m_e \alpha)$ in terms of the physical electron mass and charge, by minimization of the electron energy~\cite{Feynman}. In both cases, the spectrum scales according to variation in the selected constants, while leaving the relative internal structure of the model intact.

Na\"ive mathematical manipulations may treat the notion of the infinite cavalierly; Nature abhors it.
As the infinities of the black-body radiator led Planck to quantization of the electromagnetic field,
avoidance of Planck-scale divergences in the cosmological constant may today lead us to the No-Scale
boundary conditions, dynamically established by a suitably chosen K\"ahler potential.  The implementation of
this boundary in a manner that is consistent with precision low energy phenomenology is facilitated by
adoption of the flipped $SU(5)$ GUT structure, and the perturbing influence of extra vector-like fields.
The feasible near-term detectability of these {\it flippon} multiplets, so named for their distinctive flipped charge assignments,
presents a prime target for the ongoing LHC search.


\section{The Higgs Mass Perturbation}

In 1947, Lamb and Retherford observed the 1058~MHz splitting
of the $2 S_{1/2}$ Hydrogen level from the otherwise
identical total angular momentum doublet $2 P_{1/2}$ with unit
orbital excitation.  Coping with this small {\it finite} correction
required the budding computational apparatus of renormalization,
but the first real quantum electrodynamic calculations, attending in
isolation to the photon vacuum polarization, predicted a shift of -27~MHz
which is incorrect in both magnitude and sign.  However, the calculation was
not itself incorrect, but merely incomplete.  With the realization that
the electron propagator and vertex corrections similarly contribute
at single loop order, the amended sum agreed emphatically with experiment,
and the age of field theory was begun in earnest.  In like manner, the mild
disagreement between the baseline No-Scale \fsu5 $m_h$ prediction~\cite{Li:2011xg}
and the latest ATLAS and CMS measurements~\cite{Collaboration:2012tx,Collaboration:2012si} does not indict the validity
of the calculated model effects, so long as there remain viable cards to be played.
In fact, recognizing the potential of the {\it flippons} to elevate the Higgs mass,
but having yet no compelling experimental need to introduce the complication, we
previously favored rather heavier benchmark values for the vector-like mass $M_{\rm V}$
which would suppress their contribution~\cite{Li:2011xg}.  With the experimental
focus now becoming more clear, and our rather narrow range of bare $m_h$ predictions
appearing in this context to be somewhat light, the time has come for careful
reanalysis.  The resulting $3$-$4$~GeV upward shift in $m_h$, into the center of
the experimental limelight, strikes us as the most serendipitous of all outcomes:
small enough to justify the prior use of a first order calculation, but large enough
to be an essential final ingredient.  It provides the first concrete mechanism
for fixing the elusive $M_{\rm V}$ parameter, and directly ties the existence
of the {\it flippon} field to an immediate phenomenological consequence.
We expect this effect to decouple at leading order from the remainder of the
described phenomenology, as is the standard result in perturbative expansions.

The mechanism for the desired shift is the following pair of Yukawa
interaction terms between the MSSM Higgs and the vector-like {\it flippons}
in the superpotential, {\it cf.} Eqs.~(\ref{eq:Higgs}) and (\ref{eq:flippons}).
\begin{equation}
W = {\frac{1}{2}} Y_{xd} \, XF \, XF \, h + {\frac{1}{2}} Y_{xu} \, \overline{XF} \, \overline{XF} \, \overline{h}
\end{equation}
Being vector-like rather than chiral, the {\it flippons} are also
afforded a proper ``diagonal'' Dirac mass.
After the $SU(5)\times U(1)_X$ gauge symmetry is broken down
to the SM, the relevant Yukawa couplings are
\begin{equation}
W =  Y_{xd} XQ XD^c H_d + Y_{xu} XQ^c XD H_u~.
\end{equation}
We employ the RGE improved one-loop effective Higgs
potential approach to calculate the contributions to the lightest
CP-even Higgs boson mass from the vector-like
particles~\cite{Babu:2008ge,Martin:2009bg}.
The relevant shift in the Higgs mass-square is
approximated as~\cite{Huo:2011zt}
\begin{eqnarray}
\Delta m_h^2 &=& -\frac{N_c M_Z^2}{8\pi^2}\times \cos^22\beta~({\hat Y}_{xu}^2+{\hat Y}_{xd}^2)t_V
\nonumber \\
&+&\frac{N_cv^2}{4\pi^2}\times{\hat Y}_{xu}^4~(t_V+\frac{1}{2}X_{xu})~,
\label{Delta mhs}
\end{eqnarray}
with
\begin{eqnarray}
&{\hat Y}_{xu}=Y_{xu}\sin\beta \quad;\quad {\hat Y}_{xd}=Y_{xd}\cos\beta&
\nonumber\\
&{\tilde A}_{xu}=A_{xu}-\mu\cot\beta \quad;\quad t_V=\ln\frac{M_S^2+M_V^2}{M_V^2}&
\\
&X_{xu}=\frac{ -2M_S^2(5M_S^2+4M_V^2) + 4(3M_S^2 - 2M_V^2) {\tilde A}_{xu}^2+{\tilde A}_{xu}^4}{6(M_V^2+M_S^2)^2}&~,
\nonumber
\label{eq:vectorhiggs}
\end{eqnarray}
where $N_c$ is the number of colors carried by the vector-like fields, 
$M_S$ is the soft SUSY breaking mass evaluated at the Higgs scale,
and ${A}_{xu}$ is the soft SUSY breaking trilinear term for 
the Yukawa superpotential element $Y_{xu} XQ^c XD H_u$.
For simplicity, we take $Y_{xd}=0$.  From the two-loop 
RGE analyses, we determined that the maximal Yukawa couplings
$Y_{xu}$ are about 0.96, 1.03, and 1.0
for $\tan\beta=2$, $\tan\beta \sim 23$,  and $\tan\beta=50$, 
respectively~\cite{Huo:2011zt}, and thus
choose a working value of $Y_{xu}=1.0$.
The corrected Higgs boson mass will be 
\begin{eqnarray}
m_h ~=~ \sqrt{({\widetilde{m}}_h)^2 +\Delta m_h^2}~,~\,
\label{eq:higgs}
\end{eqnarray}
where $\widetilde{m}_h$ is the ``bare'' Higgs mass, neglecting
the shift from the vector-like coupling.
For specificity, we consider a benchmark point with inputs $M_{1/2}=518~{\rm GeV}$,
$M_V=1640$~GeV, $m_{\rm t} = 174.4$ and $\tan\beta=20.65$, which yields 
the bare Higgs mass prediction $\widetilde{m}_h=121.4$~GeV.
The SUSY spectrum for this benchmark is presented in Table~(\ref{tab:masses}).
Noting that the scalar masses will be similar to those of the top quarks,
we approximate $M_S=1$~TeV, and $A_{xu}=-1.3$~TeV.
At the benchmark, we obtain  $\Delta m_h^2=986.1~{\rm GeV}^2$, corresponding
to a corrected Higgs mass of $m_h=125.4$~GeV.

\begin{table}[ht]
  \small
    \centering
    \caption{Spectrum (in GeV) for $M_{1/2} = 518$~ GeV, $M_{V} = 1640$~GeV, $m_{t} = 174.4$~GeV, $\tan \beta$ = 20.65. Here, $\Omega_{\chi}$ = 0.1155 and the lightest neutralino is greater than 99\% Bino. For the rare process constraints, ${\rm Br} (b \to s\gamma) = 2.76 \times 10^{-4}$, $\Delta a_\mu = 12.5 \times 10^{-10}$, and Br($B^0_{s} \to \mu^+\mu^-) = 3.8 \times 10^{-9}$.	The partial lifetime for proton decay in the leading ${(e|\mu)}^{+} \pi^0 $ channels falls around $4 \times 10^{34}$~Y~\cite{Li:2010dp,Li:2010rz}.}
		\begin{tabular}{|c|c||c|c||c|c||c|c||c|c||c|c|} \hline
    $\widetilde{\chi}_{1}^{0}$&$99$&$\widetilde{\chi}_{1}^{\pm}$&$216$&$\widetilde{e}_{R}$&$196$&$\widetilde{t}_{1}$&$558$&$\widetilde{u}_{R}$&$1053$&$m_{h}$&$125.4$\\ \hline
    $\widetilde{\chi}_{2}^{0}$&$216$&$\widetilde{\chi}_{2}^{\pm}$&$900$&$\widetilde{e}_{L}$&$570$&$\widetilde{t}_{2}$&$982$&$\widetilde{u}_{L}$&$1144$&$m_{A,H}$&$972$\\ \hline
    $\widetilde{\chi}_{3}^{0}$&$896$&$\widetilde{\nu}_{e/\mu}$&$565$&$\widetilde{\tau}_{1}$&$108$&$\widetilde{b}_{1}$&$934$&$\widetilde{d}_{R}$&$1094$&$m_{H^{\pm}}$&$976$\\ \hline
    $\widetilde{\chi}_{4}^{0}$&$899$&$\widetilde{\nu}_{\tau}$&$551$&$\widetilde{\tau}_{2}$&$560$&$\widetilde{b}_{2}$&$1046$&$\widetilde{d}_{L}$&$1147$&$\widetilde{g}$&$704$\\ \hline
		\end{tabular}
		\label{tab:masses}
\end{table}

\begin{figure*}[htp]
        \centering
        \includegraphics[width=0.90\textwidth]{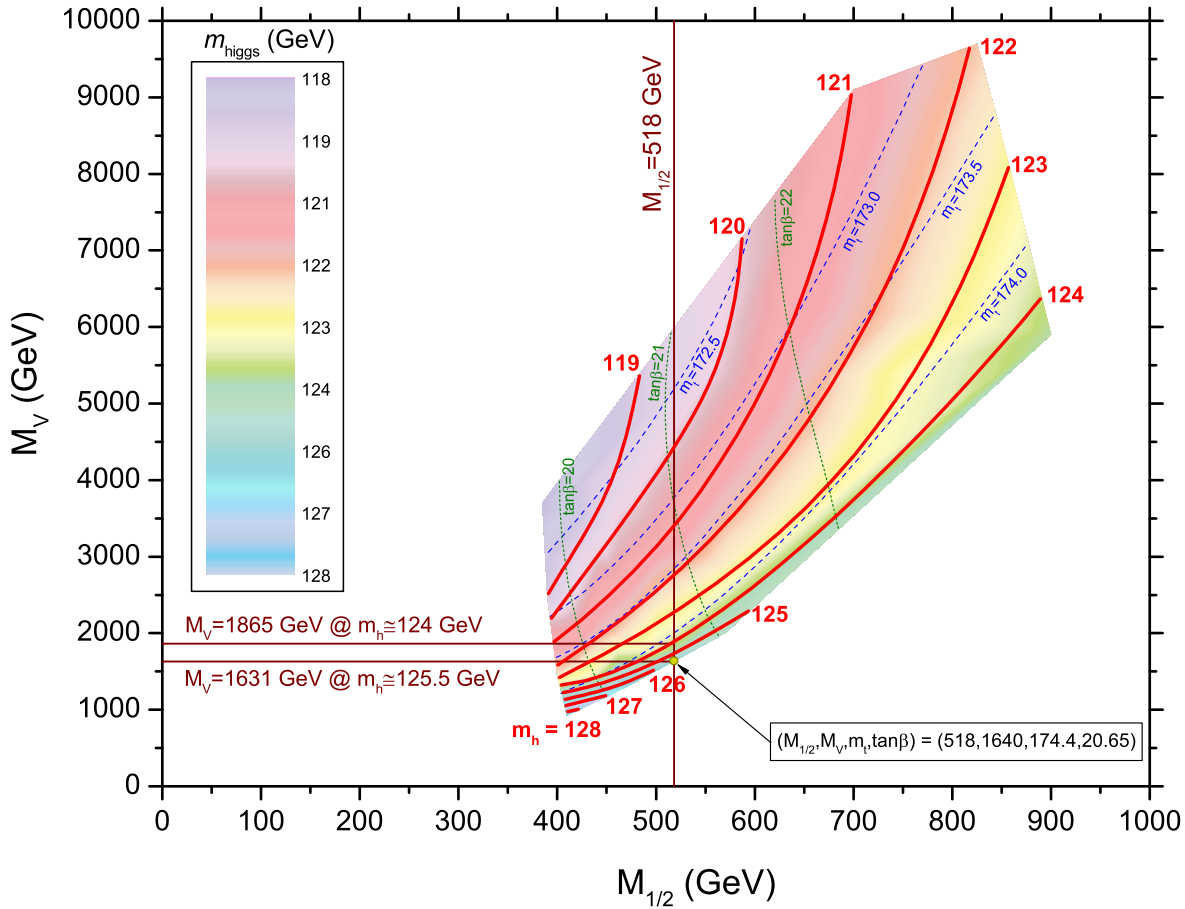}
        \caption{The space of bare-minimal constraints~\cite{Li:2011xu} on the No-Scale \fsu5 model is presented
	in the ($M_{1/2}$,$M_{\rm V}$) plane, with contour overlays designating the $\tan \beta$ and $m_{\rm t}$
	parameter ranges, in addition to the corrected Higgs mass $m_h$, inclusive of the shift from vector-like multiplet interactions.}
        \label{fig:Higgs_wedge}
\end{figure*}

\begin{figure*}[htp]
        \centering
        \includegraphics[width=0.85\textwidth]{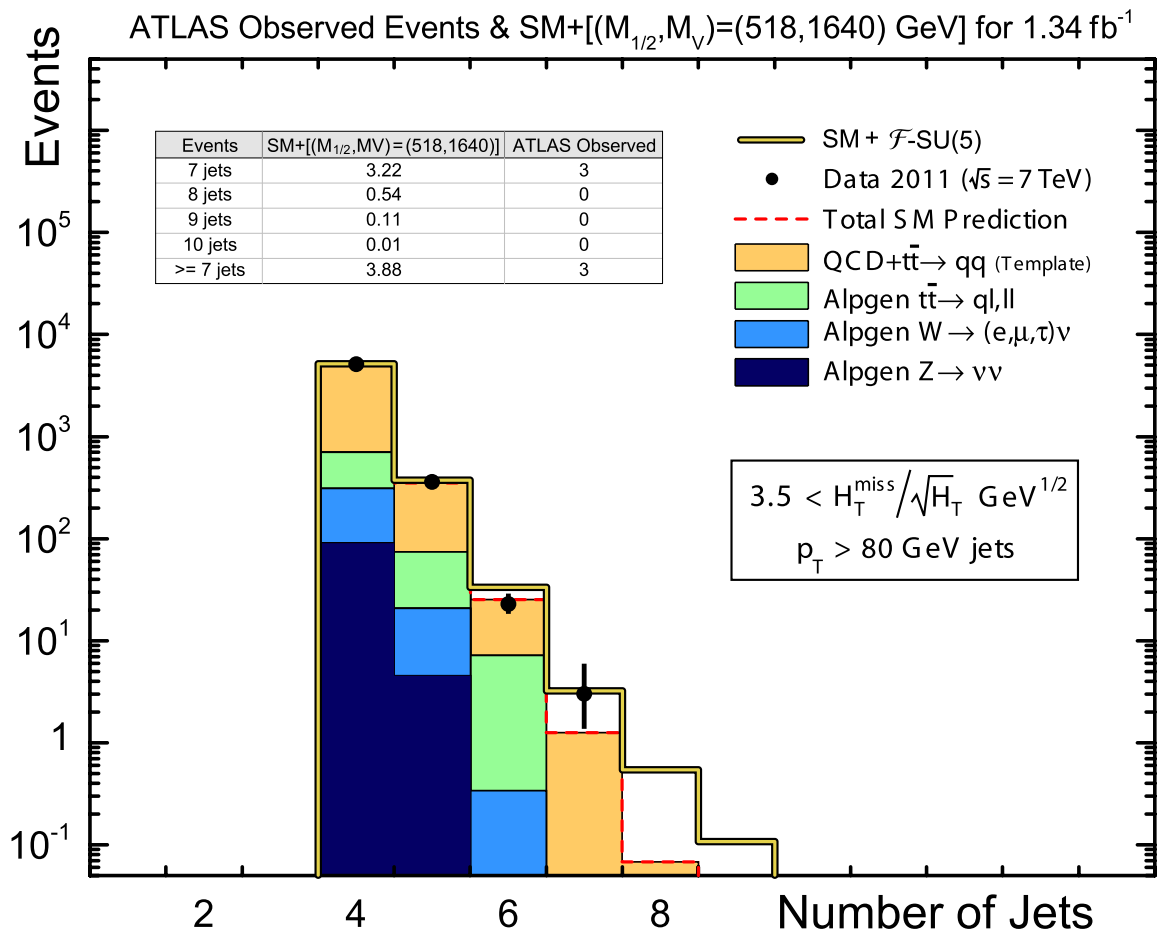}
        \caption{An ATLAS collaboration plot~\cite{Aad:2011qa} (present in the arXiv source repository supplementing
	the cited document) representing $1.34~{\rm fb}^{-1}$ of integrated luminosity
	at $\sqrt{s} = 7$~TeV is reprinted with an overlay summing our Monte Carlo collider-detector simulation of the
	No-Scale \fsu5 model benchmark ($M_{1/2}=518$~GeV, $M_{\rm V}=1640$~GeV) with the ATLAS SM background.}
        \label{fig:ATLAS_data}
\end{figure*}

\begin{figure*}[htp]
        \centering
        \includegraphics[width=0.79\textwidth]{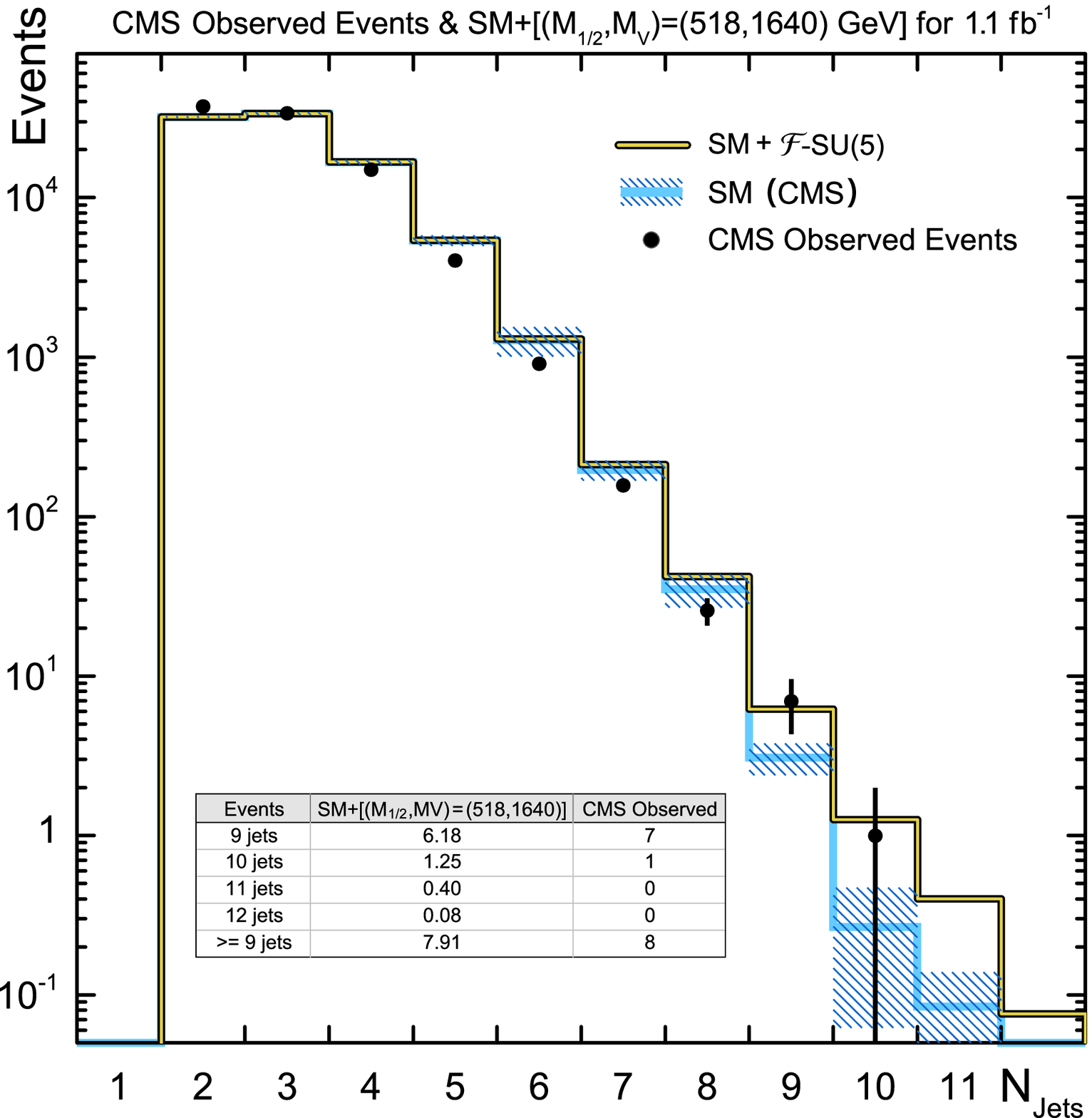}
        \caption{A CMS collaboration plot~\cite{PAS-SUS-11-003} representing $1.1~{\rm fb}^{-1}$ of integrated luminosity
	at $\sqrt{s} = 7$~TeV is reprinted with an overlay summing our Monte Carlo collider-detector simulation of the
	No-Scale \fsu5 model benchmark ($M_{1/2}=518$~GeV, $M_{\rm V}=1640$~GeV) with the CMS SM background.}
        \label{fig:CMS_data}
\end{figure*}

The parameter values at the benchmark point are not chosen idly.  In particular, the gaugino boundary mass $M_{1/2}$ is responsible
for setting the overall scale of the SUSY spectrum, and we shall demonstrate in the next section that a value in the neighborhood of $518$~GeV
is optimal for matching the observed excesses in multi-jet production events at CMS~\cite{PAS-SUS-09-001} and ATLAS~\cite{Aad:2011qa}.
By contrast, the spectrum, and thus the rate of event production, is largely indifferent to the vector-like mass.  Nevertheless,
preservation of the fundamental $B_\mu = 0$ boundary at the high scale implies an exceedingly strong parameter interdependence such
that fluctuations (for example) of the thresholds contributed by the top quark, SUSY partners, or {\it flippon} fields, might conspire against
modifications (for instance) to the Yukawa coupling boundary or $\tan \beta$, to keep the condition intact.  In Figure~(\ref{fig:Higgs_wedge}),
the contours of $\tan \beta$ are seen to run approximately along constant $M_{1/2}$, such that the jet count observations also
tightly constrain this parameter, which spans already a rather narrow range across the full parameter space.  By contrast,
the top quark mass $m_{\rm t}$ contours do vary strongly in the inverse with the vector-like mass.  We additionally overlay onto Figure~(\ref{fig:Higgs_wedge})
an extrapolation of the corrected Higgs mass.  This is achieved by use of a simplified formula for $\Delta m_h^2$~\cite{Martin:2009bg}
which implements the leading dependence of the {\it flippon} mass $M_{\rm V}$, with larger shifts corresponding to lighter vector-like multiplets.
We also account for a weaker dependence on the soft term mass by its proportionality to $M_{1/2}$, approximating $M_S = 2 M_{1/2}$.  The overall numerical
scale of the shift is calibrated against a second detailed calculation at ($M_{1/2}$,$M_V$) = (518,1840)~GeV which yields $\Delta m_h^2 = 811.5~{\rm GeV}^2$.
The extrapolation is found to agree with the original calculation at $M_V = 1640$~GeV to about 3\% in $\Delta m_h^2$.
\begin{eqnarray}
{m}_h &\simeq& \sqrt{ \widetilde{m}_h^2 + (87.81~{\rm GeV})^2 \times \left( \ln x - \frac{5}{6} +\frac{1}{x} -\frac{1}{6x^2} \right)}
\nonumber \\
x &\equiv& 1 + {\left( \frac{2 M_{1/2}}{M_{\rm V}} \right)}^2
\label{eq:m_higgs}
\end{eqnarray}
The effect of the shift term on the Higgs contours is to make the curves run more horizontally at low $M_{\rm V}$ values, in keeping with the strong
gradient in that coordinate.  This works in tandem with the top quark mass, whose elevation likewise lifts the bare Higgs mass prediction $\widetilde{m}_h$.
For both reasons, larger net values of the Higgs mass $m_h$ occur toward the lower boundary of the plot, just prior to the extremity of a 
single deviation from the top quark world average, at $m_{\rm t} \simeq 174.4$~GeV. We find that the maximum Higgs mass for $M_{1/2}$=518 GeV occurs around $M_V$=1631 GeV at $m_h \simeq$125.5 GeV, and that the Higgs mass of $m_h \simeq$124.0 GeV occurs for about $M_V$=1865 GeV, as shown in Figure~(\ref{fig:Higgs_wedge}).


\section{Next Stop-Gluino Masses}

Achieving a $125$~GeV Higgs mass in basic MSSM constructions is actually not so difficult;
it is simply a question of whether one is willing to swallow the consequences.
The bitter pill for such a heavy value of $m_h$ will be extremely heavy squarks.  To generate
$m_h=125$~ GeV will typically require squark masses in excess of 5-6 TeV, with a gluino mass of similar scale.
To achieve $m_h = 125$~GeV in this manner one may begin to contemplate soft SUSY
breaking scalar and gaugino masses so large as to imperil the very mechanism
being advanced: supersymmetry as a remedy for the ailments of the gauge hierarchy.  Of more
immediately tangible concern, gluino and squark masses approaching 5 TeV will produce no discernible
SUSY signature at the current operating energy of the LHC.  Of course, Nature cares little for such
provincial trifles; except that we believe the lightly distinctive fragrance~\cite{Li:2011av} of an
emerging SUSY signal perfuses already the early data.  A SUSY spectrum elevated to 5 TeV is far beyond the
reach of the $\sqrt{s}$=7 TeV LHC for many years to come, and could not possibly explain low-statistics signal that
may plausibly have already begun its inevitable effusion from the trillions of recorded collisions.
If the emerging SUSY signal and the putative 124--126 GeV Higgs mass are indeed legitimate, then there is another way forward. 

The modifications to the $\beta$-function coefficients from introduction of the {\it flippon} multiplets have a
parallel effect on the RGEs of the gauginos.  In particular, the color-charged gaugino mass $M_{\rm 3}$ likewise runs down flat from the
high energy boundary, obeying the relation $M_3/M_{1/2} \simeq \alpha_3(M_{\rm Z})/\alpha_3(M_{32}) \simeq \mathcal{O}\,(1)$,
which precipitates a conspicuously light gluino mass assignment.  Likewise, the large mass splitting expected from the heaviness
of the top quark via its strong coupling to the Higgs (which is also key to generating an appreciable
radiative Higgs mass shift $\Delta~m_h^2$) is responsible for a rather light stop squark $\widetilde{t}_1$.
The distinctively predictive $M({\widetilde{t_1}}) < M({\widetilde{g}}) < M({\widetilde{q}})$ mass hierarchy of a light stop
and gluino, both much lighter than all other squarks, is stable across the full No-Scale \fsu5 model space, but is
not precisely replicated in any phenomenologically favored constrained MSSM (CMSSM) constructions of which we are aware.
This spectrum generates a unique event topology starting from the pair production of heavy squarks
$\widetilde{q} \widetilde{\overline{q}}$, except for the light stop, in the initial hard scattering process,
with each squark likely to yield a quark-gluino pair $\widetilde{q} \rightarrow q \widetilde{g}$.  Each gluino may be expected
to produce events with a high multiplicity of virtual stops, via the (possibly off-shell) $\widetilde{g} \rightarrow \widetilde{t}$
transition, which in turn may terminate into hard scattering products such as $\rightarrow W^{+}W^{-} b \overline{b} \widetilde{\chi}_1^{0}$
and $W^{-} b \overline{b} \tau^{+} \nu_{\tau} \widetilde{\chi}_1^{0}$, where the $W$ bosons will produce mostly hadronic jets and some leptons.
The model described may then consistently exhibit a net product of eight or more hard jets emergent from a single squark pair production event,
passing through a single intermediate gluino pair, resulting after fragmentation in a spectacular signal of ultra-high multiplicity final state jet events. From our phenomenological analysis of the \fsu5 bare-minimally constrained parameter space in Figure~(\ref{fig:Higgs_wedge}), the gluino mass $m_{\widetilde{g}}$ ranges from about $m_{\widetilde{g}} \simeq$ 540 GeV (corresponding to an $M_{1/2} \simeq$ 385 GeV) to around $m_{\widetilde{g}} \simeq$ 1.2 TeV (corresponding to an $M_{1/2} \simeq$ 900 GeV).

We have carefully studied the expected \fsu5 production excesses in the high multiplicity jet channels~\cite{Maxin:2011hy,Li:2011gh,Li:2011rp,Li:2011fu,Li:2011av}, undertaking a detailed and comprehensive Monte Carlo simulation, employing industry standard tools~\cite{Stelzer:1994ta,MGME,Alwall:2007st,Sjostrand:2006za,PGS4}. We have painstakingly mimicked~\cite{Maxin:2011hy,Li:2011av} the leading multi-jet selection strategies of the CMS~\cite{PAS-SUS-11-003} and ATLAS~\cite{Aad:2011qa} collaborations, using a post-processing script of our own design~\cite{cutlhco}.  All 2-body SUSY processes have been included in our simulation. Our conclusion is that the best fit to the jet production excesses observed at both detectors occurs in the vicinity of the $M_{1/2} = 518$~GeV strip of Figure~(\ref{fig:Higgs_wedge}) which was isolated for attention in the prior section.  Lighter values of $M_{1/2}$ will allow for lighter {\it flippons} and a heavier top quark, and thus also a heavier Higgs.  However, values much below about $M_{1/2} = 480$ GeV are considered to be excluded for over-production of SUSY events.  Values much larger than the target range will have some difficulty achieving a sufficiently large Higgs mass.  In Figures~(\ref{fig:ATLAS_data}) and (\ref{fig:CMS_data}), we overlay counts for the No-Scale \fsu5 jet production (summed with the official SM backgrounds) onto histograms illustrating the current status of the LHC multi-jet SUSY search, representing just over $1.1~{\rm fb}^{-1}$ of luminosity integrated by the ATLAS~\cite{Aad:2011qa} and CMS~\cite{PAS-SUS-11-003} experiments, respectively.  The statistical significance of the ATLAS overproduction, as gauged by the indicator of signal (observations minus background) to background ratio $S/\sqrt{B+1}$, is quite low for $\ge$ 7 jets in the search strategy of Figure~(\ref{fig:ATLAS_data}), somewhat greater than 1.0, although the CMS overproduction significance for $\ge$ 9 jets in the search strategy of Figure~(\ref{fig:CMS_data}) is just above $2.0$. We project in Table~(\ref{tab:signals}) that the already collected $5~{\rm fb}^{-1}$ data set may be sufficient to reach the gold standard signal significance of 5, at least for the CMS search strategy, although both approaches appear to scale well with higher intensities.

\begin{table}[h]
\centering
\caption{Projections for the ATLAS and CMS signal significance at 5~${\rm fb}^{-1}$ of integrated luminosity, in the ultra-high
jet multiplicity channels.  Event counts for \fsu5 are based on our own Monte Carlo of the $M_{1/2} = 518$~GeV, $M_{V} = 1640$~GeV
benchmark.  SM backgrounds are scaled up from official collaboration estimates~\cite{PAS-SUS-11-003,Aad:2011qa}.}
\begin{tabular}{|c|c|c|c|c|c||c|c|c|c|c|}\cline{2-11}
\multicolumn{1}{c|}{} & \multicolumn{5}{|c||}{CMS~$5~{\rm fb}^{-1}$} & \multicolumn{5}{|c|}{ATLAS~$5~{\rm fb}^{-1}$} \\ \cline{2-11}
\multicolumn{1}{c|}{} &$	{\rm 9j}$&$	{\rm 10j}$&$	{\rm 11j}$&$	{\rm 12j}$&$\,{\rm \ge9j}\,$
&$	{\rm 7j}$&$	{\rm 8j}$&$	{\rm 9j}$&$	{\rm 10j}$&$\,{\rm \ge7j}\,$ \\ \hline	\hline
\fsu5 &$	14.0$&$	4.5$&$	1.4$&$	0.3$&$	20.3$&$	7.3$&$	1.8$&$	0.4$&$	0.1$&$	9.6$ \\ \hline
${\rm SM}	$&$	14.0$&$	1.2$&$	0.4$&$	0.0$&$	15.6$&$	4.7$&$	0.3$&$	0.0$&$	0.0$&$	4.9$ \\ \hline \hline
$S/ \sqrt{B+1}$&$	3.6$&$	3.0$&$	1.2$&$	0.3$&${{\bf 5.0}}$&$	3.1$&$	1.6$&$	0.4$&$	0.1$&${{\bf 3.9}}$ \\ \hline
\end{tabular}
\label{tab:signals}
\end{table}


\section{Conclusions}

We have described a model named No-Scale \fsu5 which is simultaneously capable of explaining the dual signals emerging
at the LHC of i) a 124--126 GeV Higgs boson mass $m_h$, and ii) tantalizing low-statistics excesses in the multi-jet data
which may attributable to supersymmetry.  These targets tend to be mutually exclusive in more conventional approaches.
The unified mechanism responsible for both effects is the introduction of a rather unique set of
vector-like multiplets at the TeV scale, dubbed {\it flippons}, which i) can elevate $m_h$ by around $3-4$~GeV
via radiative loop corrections, and ii) flatten the running of the strong coupling and color-charged gaugino, resulting
in a prominent collider signal from production of light gluino pairs.  This well motivated theoretical framework maintains consistency
with all key phenomenological constraints, and all residual parameterization freedom may in principle be fixed by a
combination of the two experiments described.
We project that the already collected luminosity of $5~{\rm fb}^{-1}$ may be sufficient to definitively
establish the status of this model, given appropriate data selection cuts.


\begin{acknowledgments}
We acknowledge the generous contribution of Chunli Tong and Yunjie Huo of the
State Key Laboratory of Theoretical Physics in Beijing for
their calculation of Eqs.~(\ref{Delta mhs}-\ref{eq:higgs}).
This research was supported in part 
by the DOE grant DE-FG03-95-Er-40917 (TL and DVN),
by the Natural Science Foundation of China 
under grant numbers 10821504 and 11075194 (TL),
by the Mitchell-Heep Chair in High Energy Physics (JAM),
and by the Sam Houston State University
2011 Enhancement Research Grant program (JWW).
We also thank Sam Houston State University
for providing high performance computing resources.  
\end{acknowledgments}


\bibliography{bibliography}

\begin{thebibliography}{59}
\expandafter\ifx\csname natexlab\endcsname\relax\def\natexlab#1{#1}\fi
\expandafter\ifx\csname bibnamefont\endcsname\relax
  \def\bibnamefont#1{#1}\fi
\expandafter\ifx\csname bibfnamefont\endcsname\relax
  \def\bibfnamefont#1{#1}\fi
\expandafter\ifx\csname citenamefont\endcsname\relax
  \def\citenamefont#1{#1}\fi
\expandafter\ifx\csname url\endcsname\relax
  \def\url#1{\texttt{#1}}\fi
\expandafter\ifx\csname urlprefix\endcsname\relax\def\urlprefix{URL }\fi
\providecommand{\bibinfo}[2]{#2}
\providecommand{\eprint}[2][]{\url{#2}}

\bibitem[{\citenamefont{Georgi et~al.}(1978)\citenamefont{Georgi, Glashow,
  Machacek, and Nanopoulos}}]{Georgi:1977gs}
\bibinfo{author}{\bibfnamefont{H.}~\bibnamefont{Georgi}},
  \bibinfo{author}{\bibfnamefont{S.}~\bibnamefont{Glashow}},
  \bibinfo{author}{\bibfnamefont{M.}~\bibnamefont{Machacek}}, \bibnamefont{and}
  \bibinfo{author}{\bibfnamefont{D.~V.} \bibnamefont{Nanopoulos}},
  {``}\bibinfo{title}{{Higgs Bosons from Two Gluon Annihilation in Proton
  Proton Collisions}},{''} \bibinfo{journal}{Phys.Rev.Lett.}
  \textbf{\bibinfo{volume}{40}}, \bibinfo{pages}{692} (\bibinfo{year}{1978}).

\bibitem[{\citenamefont{Ellis et~al.}(1976)\citenamefont{Ellis, Gaillard, and
  Nanopoulos}}]{Ellis:1975ap}
\bibinfo{author}{\bibfnamefont{J.~R.} \bibnamefont{Ellis}},
  \bibinfo{author}{\bibfnamefont{M.~K.} \bibnamefont{Gaillard}},
  \bibnamefont{and} \bibinfo{author}{\bibfnamefont{D.~V.}
  \bibnamefont{Nanopoulos}}, {``}\bibinfo{title}{{A Phenomenological Profile of
  the Higgs Boson}},{''} \bibinfo{journal}{Nucl.Phys.}
  \textbf{\bibinfo{volume}{B106}}, \bibinfo{pages}{292} (\bibinfo{year}{1976}).

\bibitem[{\citenamefont{Li et~al.}(2011{\natexlab{a}})\citenamefont{Li, Maxin,
  Nanopoulos, and Walker}}]{Li:2010ws}
\bibinfo{author}{\bibfnamefont{T.}~\bibnamefont{Li}},
  \bibinfo{author}{\bibfnamefont{J.~A.} \bibnamefont{Maxin}},
  \bibinfo{author}{\bibfnamefont{D.~V.} \bibnamefont{Nanopoulos}},
  \bibnamefont{and} \bibinfo{author}{\bibfnamefont{J.~W.}
  \bibnamefont{Walker}}, {``}\bibinfo{title}{{The Golden Point of No-Scale and
  No-Parameter ${\cal F}$-$SU(5)$}},{''} \bibinfo{journal}{Phys. Rev.}
  \textbf{\bibinfo{volume}{D83}}, \bibinfo{pages}{056015}
  (\bibinfo{year}{2011}{\natexlab{a}}), \eprint{1007.5100}.

\bibitem[{\citenamefont{Li et~al.}(2011{\natexlab{b}})\citenamefont{Li, Maxin,
  Nanopoulos, and Walker}}]{Li:2010mi}
\bibinfo{author}{\bibfnamefont{T.}~\bibnamefont{Li}},
  \bibinfo{author}{\bibfnamefont{J.~A.} \bibnamefont{Maxin}},
  \bibinfo{author}{\bibfnamefont{D.~V.} \bibnamefont{Nanopoulos}},
  \bibnamefont{and} \bibinfo{author}{\bibfnamefont{J.~W.}
  \bibnamefont{Walker}}, {``}\bibinfo{title}{{The Golden Strip of Correlated
  Top Quark, Gaugino, and Vectorlike Mass In No-Scale, No-Parameter
  F-SU(5)}},{''} \bibinfo{journal}{Phys. Lett.}
  \textbf{\bibinfo{volume}{B699}}, \bibinfo{pages}{164}
  (\bibinfo{year}{2011}{\natexlab{b}}), \eprint{1009.2981}.

\bibitem[{\citenamefont{Li et~al.}(2011{\natexlab{c}})\citenamefont{Li, Maxin,
  Nanopoulos, and Walker}}]{Li:2010uu}
\bibinfo{author}{\bibfnamefont{T.}~\bibnamefont{Li}},
  \bibinfo{author}{\bibfnamefont{J.~A.} \bibnamefont{Maxin}},
  \bibinfo{author}{\bibfnamefont{D.~V.} \bibnamefont{Nanopoulos}},
  \bibnamefont{and} \bibinfo{author}{\bibfnamefont{J.~W.}
  \bibnamefont{Walker}}, {``}\bibinfo{title}{{Super No-Scale ${\cal
  F}$-$SU(5)$: Resolving the Gauge Hierarchy Problem by Dynamic Determination
  of $M_{1/2}$ and $\tan\beta$}},{''} \bibinfo{journal}{Phys. Lett. B}
  \textbf{\bibinfo{volume}{703}}, \bibinfo{pages}{469}
  (\bibinfo{year}{2011}{\natexlab{c}}), \eprint{1010.4550}.

\bibitem[{\citenamefont{Li et~al.}(2011{\natexlab{d}})\citenamefont{Li, Maxin,
  Nanopoulos, and Walker}}]{Li:2011dw}
\bibinfo{author}{\bibfnamefont{T.}~\bibnamefont{Li}},
  \bibinfo{author}{\bibfnamefont{J.~A.} \bibnamefont{Maxin}},
  \bibinfo{author}{\bibfnamefont{D.~V.} \bibnamefont{Nanopoulos}},
  \bibnamefont{and} \bibinfo{author}{\bibfnamefont{J.~W.}
  \bibnamefont{Walker}}, {``}\bibinfo{title}{{Blueprints of the No-Scale
  Multiverse at the LHC}},{''} \bibinfo{journal}{Phys. Rev.}
  \textbf{\bibinfo{volume}{D84}}, \bibinfo{pages}{056016}
  (\bibinfo{year}{2011}{\natexlab{d}}), \eprint{1101.2197}.

\bibitem[{\citenamefont{Li et~al.}(2011{\natexlab{e}})\citenamefont{Li, Maxin,
  Nanopoulos, and Walker}}]{Li:2011hr}
\bibinfo{author}{\bibfnamefont{T.}~\bibnamefont{Li}},
  \bibinfo{author}{\bibfnamefont{J.~A.} \bibnamefont{Maxin}},
  \bibinfo{author}{\bibfnamefont{D.~V.} \bibnamefont{Nanopoulos}},
  \bibnamefont{and} \bibinfo{author}{\bibfnamefont{J.~W.}
  \bibnamefont{Walker}}, {``}\bibinfo{title}{{Ultra High Jet Signals from
  Stringy No-Scale Supergravity}},{''} (\bibinfo{year}{2011}{\natexlab{e}}),
  \eprint{1103.2362}.

\bibitem[{\citenamefont{Li et~al.}(2011{\natexlab{f}})\citenamefont{Li, Maxin,
  Nanopoulos, and Walker}}]{Maxin:2011hy}
\bibinfo{author}{\bibfnamefont{T.}~\bibnamefont{Li}},
  \bibinfo{author}{\bibfnamefont{J.~A.} \bibnamefont{Maxin}},
  \bibinfo{author}{\bibfnamefont{D.~V.} \bibnamefont{Nanopoulos}},
  \bibnamefont{and} \bibinfo{author}{\bibfnamefont{J.~W.}
  \bibnamefont{Walker}}, {``}\bibinfo{title}{{The Ultrahigh jet multiplicity
  signal of stringy no-scale $\cal{F}$-$SU(5)$ at the $\sqrt{s}= 7$ TeV
  LHC}},{''} \bibinfo{journal}{Phys.Rev.} \textbf{\bibinfo{volume}{D84}},
  \bibinfo{pages}{076003} (\bibinfo{year}{2011}{\natexlab{f}}),
  \eprint{1103.4160}.

\bibitem[{\citenamefont{Li et~al.}(2012{\natexlab{a}})\citenamefont{Li, Maxin,
  Nanopoulos, and Walker}}]{Li:2011xu}
\bibinfo{author}{\bibfnamefont{T.}~\bibnamefont{Li}},
  \bibinfo{author}{\bibfnamefont{J.~A.} \bibnamefont{Maxin}},
  \bibinfo{author}{\bibfnamefont{D.~V.} \bibnamefont{Nanopoulos}},
  \bibnamefont{and} \bibinfo{author}{\bibfnamefont{J.~W.}
  \bibnamefont{Walker}}, {``}\bibinfo{title}{{The Unification of Dynamical
  Determination and Bare Minimal Phenomenological Constraints in No-Scale
  \cal{F}- SU(5)}},{''} \bibinfo{journal}{Phys.Rev.} \textbf{\bibinfo{volume}{D
  In Press}} (\bibinfo{year}{2012}{\natexlab{a}}), \eprint{1105.3988}.

\bibitem[{\citenamefont{Li et~al.}(2011{\natexlab{g}})\citenamefont{Li, Maxin,
  Nanopoulos, and Walker}}]{Li:2011in}
\bibinfo{author}{\bibfnamefont{T.}~\bibnamefont{Li}},
  \bibinfo{author}{\bibfnamefont{J.~A.} \bibnamefont{Maxin}},
  \bibinfo{author}{\bibfnamefont{D.~V.} \bibnamefont{Nanopoulos}},
  \bibnamefont{and} \bibinfo{author}{\bibfnamefont{J.~W.}
  \bibnamefont{Walker}}, {``}\bibinfo{title}{{The Race for Supersymmetric Dark
  Matter at XENON100 and the LHC: Stringy Correlations from No-Scale
  \cal{F}-SU(5)}},{''} (\bibinfo{year}{2011}{\natexlab{g}}),
  \eprint{1106.1165}.

\bibitem[{\citenamefont{Li et~al.}(2012{\natexlab{b}})\citenamefont{Li, Maxin,
  Nanopoulos, and Walker}}]{Li:2011gh}
\bibinfo{author}{\bibfnamefont{T.}~\bibnamefont{Li}},
  \bibinfo{author}{\bibfnamefont{J.~A.} \bibnamefont{Maxin}},
  \bibinfo{author}{\bibfnamefont{D.~V.} \bibnamefont{Nanopoulos}},
  \bibnamefont{and} \bibinfo{author}{\bibfnamefont{J.~W.}
  \bibnamefont{Walker}}, {``}\bibinfo{title}{{A Two-Tiered Correlation of Dark
  Matter with Missing Transverse Energy: Reconstructing the Lightest
  Supersymmetric Particle Mass at the LHC}},{''} \bibinfo{journal}{JHEP}
  \textbf{\bibinfo{volume}{In Press}} (\bibinfo{year}{2012}{\natexlab{b}}),
  \eprint{1107.2375}.

\bibitem[{\citenamefont{Li et~al.}(2012{\natexlab{c}})\citenamefont{Li, Maxin,
  Nanopoulos, and Walker}}]{Li:2011rp}
\bibinfo{author}{\bibfnamefont{T.}~\bibnamefont{Li}},
  \bibinfo{author}{\bibfnamefont{J.~A.} \bibnamefont{Maxin}},
  \bibinfo{author}{\bibfnamefont{D.~V.} \bibnamefont{Nanopoulos}},
  \bibnamefont{and} \bibinfo{author}{\bibfnamefont{J.~W.}
  \bibnamefont{Walker}}, {``}\bibinfo{title}{{Prospects for Discovery of
  Supersymmetric No-Scale F-SU(5) at The Once and Future LHC}},{''}
  \bibinfo{journal}{Nucl.Phys.} \textbf{\bibinfo{volume}{B859}},
  \bibinfo{pages}{96} (\bibinfo{year}{2012}{\natexlab{c}}), \eprint{1107.3825}.

\bibitem[{\citenamefont{Li et~al.}(2011{\natexlab{h}})\citenamefont{Li, Maxin,
  Nanopoulos, and Walker}}]{Li:2011fu}
\bibinfo{author}{\bibfnamefont{T.}~\bibnamefont{Li}},
  \bibinfo{author}{\bibfnamefont{J.~A.} \bibnamefont{Maxin}},
  \bibinfo{author}{\bibfnamefont{D.~V.} \bibnamefont{Nanopoulos}},
  \bibnamefont{and} \bibinfo{author}{\bibfnamefont{J.~W.}
  \bibnamefont{Walker}}, {``}\bibinfo{title}{{Has SUSY Gone Undetected in 9-jet
  Events? A Ten-Fold Enhancement in the LHC Signal Efficiency}},{''}
  (\bibinfo{year}{2011}{\natexlab{h}}), \eprint{1108.5169}.

\bibitem[{\citenamefont{Li et~al.}(2011{\natexlab{i}})\citenamefont{Li, Maxin,
  Nanopoulos, and Walker}}]{Li:2011ex}
\bibinfo{author}{\bibfnamefont{T.}~\bibnamefont{Li}},
  \bibinfo{author}{\bibfnamefont{J.~A.} \bibnamefont{Maxin}},
  \bibinfo{author}{\bibfnamefont{D.~V.} \bibnamefont{Nanopoulos}},
  \bibnamefont{and} \bibinfo{author}{\bibfnamefont{J.~W.}
  \bibnamefont{Walker}}, {``}\bibinfo{title}{{The F-Landscape: Dynamically
  Determining the Multiverse}},{''} (\bibinfo{year}{2011}{\natexlab{i}}),
  \eprint{1111.0236}.

\bibitem[{\citenamefont{Li et~al.}(2011{\natexlab{j}})\citenamefont{Li, Maxin,
  Nanopoulos, and Walker}}]{Li:2011av}
\bibinfo{author}{\bibfnamefont{T.}~\bibnamefont{Li}},
  \bibinfo{author}{\bibfnamefont{J.~A.} \bibnamefont{Maxin}},
  \bibinfo{author}{\bibfnamefont{D.~V.} \bibnamefont{Nanopoulos}},
  \bibnamefont{and} \bibinfo{author}{\bibfnamefont{J.~W.}
  \bibnamefont{Walker}}, {``}\bibinfo{title}{{Profumo di SUSY: Suggestive
  Correlations in the ATLAS and CMS High Jet Multiplicity Data}},{''}
  (\bibinfo{year}{2011}{\natexlab{j}}), \eprint{1111.4204}.

\bibitem[{\citenamefont{Barr}(1982)}]{Barr:1981qv}
\bibinfo{author}{\bibfnamefont{S.~M.} \bibnamefont{Barr}},
  {``}\bibinfo{title}{{A New Symmetry Breaking Pattern for $SO(10)$ and Proton
  Decay}},{''} \bibinfo{journal}{Phys. Lett.} \textbf{\bibinfo{volume}{B112}},
  \bibinfo{pages}{219} (\bibinfo{year}{1982}).

\bibitem[{\citenamefont{Derendinger et~al.}(1984)\citenamefont{Derendinger,
  Kim, and Nanopoulos}}]{Derendinger:1983aj}
\bibinfo{author}{\bibfnamefont{J.~P.} \bibnamefont{Derendinger}},
  \bibinfo{author}{\bibfnamefont{J.~E.} \bibnamefont{Kim}}, \bibnamefont{and}
  \bibinfo{author}{\bibfnamefont{D.~V.} \bibnamefont{Nanopoulos}},
  {``}\bibinfo{title}{{Anti-$SU(5)$}},{''} \bibinfo{journal}{Phys. Lett.}
  \textbf{\bibinfo{volume}{B139}}, \bibinfo{pages}{170} (\bibinfo{year}{1984}).

\bibitem[{\citenamefont{Antoniadis et~al.}(1987)\citenamefont{Antoniadis,
  Ellis, Hagelin, and Nanopoulos}}]{Antoniadis:1987dx}
\bibinfo{author}{\bibfnamefont{I.}~\bibnamefont{Antoniadis}},
  \bibinfo{author}{\bibfnamefont{J.~R.} \bibnamefont{Ellis}},
  \bibinfo{author}{\bibfnamefont{J.~S.} \bibnamefont{Hagelin}},
  \bibnamefont{and} \bibinfo{author}{\bibfnamefont{D.~V.}
  \bibnamefont{Nanopoulos}}, {``}\bibinfo{title}{{Supersymmetric Flipped
  $SU(5)$ Revitalized}},{''} \bibinfo{journal}{Phys. Lett.}
  \textbf{\bibinfo{volume}{B194}}, \bibinfo{pages}{231} (\bibinfo{year}{1987}).

\bibitem[{\citenamefont{Jiang et~al.}(2007)\citenamefont{Jiang, Li, and
  Nanopoulos}}]{Jiang:2006hf}
\bibinfo{author}{\bibfnamefont{J.}~\bibnamefont{Jiang}},
  \bibinfo{author}{\bibfnamefont{T.}~\bibnamefont{Li}}, \bibnamefont{and}
  \bibinfo{author}{\bibfnamefont{D.~V.} \bibnamefont{Nanopoulos}},
  {``}\bibinfo{title}{{Testable Flipped $SU(5) \times U(1)_X$ Models}},{''}
  \bibinfo{journal}{Nucl. Phys.} \textbf{\bibinfo{volume}{B772}},
  \bibinfo{pages}{49} (\bibinfo{year}{2007}), \eprint{hep-ph/0610054}.

\bibitem[{\citenamefont{Jiang et~al.}(2009)\citenamefont{Jiang, Li, Nanopoulos,
  and Xie}}]{Jiang:2009zza}
\bibinfo{author}{\bibfnamefont{J.}~\bibnamefont{Jiang}},
  \bibinfo{author}{\bibfnamefont{T.}~\bibnamefont{Li}},
  \bibinfo{author}{\bibfnamefont{D.~V.} \bibnamefont{Nanopoulos}},
  \bibnamefont{and} \bibinfo{author}{\bibfnamefont{D.}~\bibnamefont{Xie}},
  {``}\bibinfo{title}{{F-$SU(5)$}},{''} \bibinfo{journal}{Phys. Lett.}
  \textbf{\bibinfo{volume}{B677}}, \bibinfo{pages}{322} (\bibinfo{year}{2009}).

\bibitem[{\citenamefont{Jiang et~al.}(2010)\citenamefont{Jiang, Li, Nanopoulos,
  and Xie}}]{Jiang:2009za}
\bibinfo{author}{\bibfnamefont{J.}~\bibnamefont{Jiang}},
  \bibinfo{author}{\bibfnamefont{T.}~\bibnamefont{Li}},
  \bibinfo{author}{\bibfnamefont{D.~V.} \bibnamefont{Nanopoulos}},
  \bibnamefont{and} \bibinfo{author}{\bibfnamefont{D.}~\bibnamefont{Xie}},
  {``}\bibinfo{title}{{Flipped $SU(5) \times U(1)_X$ Models from
  F-Theory}},{''} \bibinfo{journal}{Nucl. Phys.}
  \textbf{\bibinfo{volume}{B830}}, \bibinfo{pages}{195} (\bibinfo{year}{2010}),
  \eprint{0905.3394}.

\bibitem[{\citenamefont{Li et~al.}(2011{\natexlab{k}})\citenamefont{Li,
  Nanopoulos, and Walker}}]{Li:2010dp}
\bibinfo{author}{\bibfnamefont{T.}~\bibnamefont{Li}},
  \bibinfo{author}{\bibfnamefont{D.~V.} \bibnamefont{Nanopoulos}},
  \bibnamefont{and} \bibinfo{author}{\bibfnamefont{J.~W.}
  \bibnamefont{Walker}}, {``}\bibinfo{title}{{Elements of F-ast Proton
  Decay}},{''} \bibinfo{journal}{Nucl. Phys.} \textbf{\bibinfo{volume}{B846}},
  \bibinfo{pages}{43} (\bibinfo{year}{2011}{\natexlab{k}}), \eprint{1003.2570}.

\bibitem[{\citenamefont{Li et~al.}(2011{\natexlab{l}})\citenamefont{Li, Maxin,
  Nanopoulos, and Walker}}]{Li:2010rz}
\bibinfo{author}{\bibfnamefont{T.}~\bibnamefont{Li}},
  \bibinfo{author}{\bibfnamefont{J.~A.} \bibnamefont{Maxin}},
  \bibinfo{author}{\bibfnamefont{D.~V.} \bibnamefont{Nanopoulos}},
  \bibnamefont{and} \bibinfo{author}{\bibfnamefont{J.~W.}
  \bibnamefont{Walker}}, {``}\bibinfo{title}{{Dark Matter, Proton Decay and
  Other Phenomenological Constraints in ${\cal F}$-SU(5)}},{''}
  \bibinfo{journal}{Nucl.Phys.} \textbf{\bibinfo{volume}{B848}},
  \bibinfo{pages}{314} (\bibinfo{year}{2011}{\natexlab{l}}),
  \eprint{1003.4186}.

\bibitem[{\citenamefont{Cremmer et~al.}(1983)\citenamefont{Cremmer, Ferrara,
  Kounnas, and Nanopoulos}}]{Cremmer:1983bf}
\bibinfo{author}{\bibfnamefont{E.}~\bibnamefont{Cremmer}},
  \bibinfo{author}{\bibfnamefont{S.}~\bibnamefont{Ferrara}},
  \bibinfo{author}{\bibfnamefont{C.}~\bibnamefont{Kounnas}}, \bibnamefont{and}
  \bibinfo{author}{\bibfnamefont{D.~V.} \bibnamefont{Nanopoulos}},
  {``}\bibinfo{title}{{Naturally Vanishing Cosmological Constant in $N=1$
  Supergravity}},{''} \bibinfo{journal}{Phys. Lett.}
  \textbf{\bibinfo{volume}{B133}}, \bibinfo{pages}{61} (\bibinfo{year}{1983}).

\bibitem[{\citenamefont{Ellis et~al.}(1984{\natexlab{a}})\citenamefont{Ellis,
  Lahanas, Nanopoulos, and Tamvakis}}]{Ellis:1983sf}
\bibinfo{author}{\bibfnamefont{J.~R.} \bibnamefont{Ellis}},
  \bibinfo{author}{\bibfnamefont{A.~B.} \bibnamefont{Lahanas}},
  \bibinfo{author}{\bibfnamefont{D.~V.} \bibnamefont{Nanopoulos}},
  \bibnamefont{and} \bibinfo{author}{\bibfnamefont{K.}~\bibnamefont{Tamvakis}},
  {``}\bibinfo{title}{{No-Scale Supersymmetric Standard Model}},{''}
  \bibinfo{journal}{Phys. Lett.} \textbf{\bibinfo{volume}{B134}},
  \bibinfo{pages}{429} (\bibinfo{year}{1984}{\natexlab{a}}).

\bibitem[{\citenamefont{Ellis et~al.}(1984{\natexlab{b}})\citenamefont{Ellis,
  Kounnas, and Nanopoulos}}]{Ellis:1983ei}
\bibinfo{author}{\bibfnamefont{J.~R.} \bibnamefont{Ellis}},
  \bibinfo{author}{\bibfnamefont{C.}~\bibnamefont{Kounnas}}, \bibnamefont{and}
  \bibinfo{author}{\bibfnamefont{D.~V.} \bibnamefont{Nanopoulos}},
  {``}\bibinfo{title}{{Phenomenological $SU(1,1)$ Supergravity}},{''}
  \bibinfo{journal}{Nucl. Phys.} \textbf{\bibinfo{volume}{B241}},
  \bibinfo{pages}{406} (\bibinfo{year}{1984}{\natexlab{b}}).

\bibitem[{\citenamefont{Ellis et~al.}(1984{\natexlab{c}})\citenamefont{Ellis,
  Kounnas, and Nanopoulos}}]{Ellis:1984bm}
\bibinfo{author}{\bibfnamefont{J.~R.} \bibnamefont{Ellis}},
  \bibinfo{author}{\bibfnamefont{C.}~\bibnamefont{Kounnas}}, \bibnamefont{and}
  \bibinfo{author}{\bibfnamefont{D.~V.} \bibnamefont{Nanopoulos}},
  {``}\bibinfo{title}{{No Scale Supersymmetric Guts}},{''}
  \bibinfo{journal}{Nucl. Phys.} \textbf{\bibinfo{volume}{B247}},
  \bibinfo{pages}{373} (\bibinfo{year}{1984}{\natexlab{c}}).

\bibitem[{\citenamefont{Lahanas and Nanopoulos}(1987)}]{Lahanas:1986uc}
\bibinfo{author}{\bibfnamefont{A.~B.} \bibnamefont{Lahanas}} \bibnamefont{and}
  \bibinfo{author}{\bibfnamefont{D.~V.} \bibnamefont{Nanopoulos}},
  {``}\bibinfo{title}{{The Road to No Scale Supergravity}},{''}
  \bibinfo{journal}{Phys. Rept.} \textbf{\bibinfo{volume}{145}},
  \bibinfo{pages}{1} (\bibinfo{year}{1987}).

\bibitem[{\citenamefont{Li et~al.}(2012{\natexlab{d}})\citenamefont{Li, Maxin,
  Nanopoulos, and Walker}}]{Li:2011xg}
\bibinfo{author}{\bibfnamefont{T.}~\bibnamefont{Li}},
  \bibinfo{author}{\bibfnamefont{J.~A.} \bibnamefont{Maxin}},
  \bibinfo{author}{\bibfnamefont{D.~V.} \bibnamefont{Nanopoulos}},
  \bibnamefont{and} \bibinfo{author}{\bibfnamefont{J.~W.}
  \bibnamefont{Walker}}, {``}\bibinfo{title}{{Natural Predictions for the Higgs
  Boson Mass and Supersymmetric Contributions to Rare Processes}},{''}
  \bibinfo{journal}{Phys.Lett.} \textbf{\bibinfo{volume}{B708}},
  \bibinfo{pages}{93} (\bibinfo{year}{2012}{\natexlab{d}}), \eprint{1109.2110}.

\bibitem[{\citenamefont{CMS}(2011)}]{PAS-HIG-11-022}
\bibinfo{author}{\bibnamefont{CMS}}, {``}\bibinfo{title}{{Search for standard
  model Higgs boson in pp collisions at $\sqrt{s}$ = 7 TeV and integrated
  luminosity up to 1.7 $fb^{-1}$}},{''} (\bibinfo{year}{2011}),
  \bibinfo{note}{{CMS-PAS-HIG-11-022}}, \urlprefix\url{http://cdsweb.cern.ch/}.

\bibitem[{\citenamefont{ATLAS}(2011{\natexlab{a}})}]{ATLAS-CONF-135}
\bibinfo{author}{\bibnamefont{ATLAS}}, {``}\bibinfo{title}{{Update of the
  Combination of Higgs Boson Searches in 1.0 to 2.3 $fb^{-1}$ of pp Collisions
  Data Taken at $\sqrt{s}$ = 7 TeV with the ATLAS Experiment at the LHC}},{''}
  (\bibinfo{year}{2011}{\natexlab{a}}), \bibinfo{note}{{ATLAS-CONF-2011-135}},
  \urlprefix\url{https://atlas.web.cern.ch/}.

\bibitem[{\citenamefont{ATLAS}(2011{\natexlab{b}})}]{ATLAS:2011ww}
\bibinfo{author}{\bibnamefont{ATLAS}}, {``}\bibinfo{title}{{Search for the
  Standard Model Higgs boson in the two photon decay channel with the ATLAS
  detector at the LHC}},{''} (\bibinfo{year}{2011}{\natexlab{b}}),
  \eprint{1108.5895}.

\bibitem[{\citenamefont{CDF/D0}(2011)}]{:2011ra}
\bibinfo{author}{\bibnamefont{CDF/D0}}, {``}\bibinfo{title}{{Combined CDF and
  D0 Searches for the Standard Model Higgs Boson Decaying to Two Photons with
  up to 8.2 $fb^{-1}$}},{''} (\bibinfo{year}{2011}), \eprint{1107.4960}.

\bibitem[{\citenamefont{CMS}(2012)}]{Collaboration:2012tx}
\bibinfo{author}{\bibnamefont{CMS}}, {``}\bibinfo{title}{{Combined results of
  searches for the standard model Higgs boson in pp collisions at $\sqrt{s}$ =
  7 TeV}},{''} (\bibinfo{year}{2012}), \eprint{1202.1488}.

\bibitem[{\citenamefont{ATLAS}(2012)}]{Collaboration:2012si}
\bibinfo{author}{\bibnamefont{ATLAS}}, {``}\bibinfo{title}{{Combined search for
  the Standard Model Higgs boson using up to 4.9 $fb^{-1}$ of pp collision data
  at $\sqrt{s}$ = 7 TeV with the ATLAS detector at the LHC}},{''}
  (\bibinfo{year}{2012}), \eprint{1202.1408}.

\bibitem[{\citenamefont{Moroi and Okada}(1992)}]{Moroi:1992zk}
\bibinfo{author}{\bibfnamefont{T.}~\bibnamefont{Moroi}} \bibnamefont{and}
  \bibinfo{author}{\bibfnamefont{Y.}~\bibnamefont{Okada}},
  {``}\bibinfo{title}{{Upper bound of the lightest neutral Higgs mass in
  extended supersymmetric Standard Models}},{''} \bibinfo{journal}{Phys.Lett.}
  \textbf{\bibinfo{volume}{B295}}, \bibinfo{pages}{73} (\bibinfo{year}{1992}).

\bibitem[{\citenamefont{Babu et~al.}(2008)\citenamefont{Babu, Gogoladze,
  Rehman, and Shafi}}]{Babu:2008ge}
\bibinfo{author}{\bibfnamefont{K.}~\bibnamefont{Babu}},
  \bibinfo{author}{\bibfnamefont{I.}~\bibnamefont{Gogoladze}},
  \bibinfo{author}{\bibfnamefont{M.~U.} \bibnamefont{Rehman}},
  \bibnamefont{and} \bibinfo{author}{\bibfnamefont{Q.}~\bibnamefont{Shafi}},
  {``}\bibinfo{title}{{Higgs Boson Mass, Sparticle Spectrum and Little
  Hierarchy Problem in Extended MSSM}},{''} \bibinfo{journal}{Phys.Rev.}
  \textbf{\bibinfo{volume}{D78}}, \bibinfo{pages}{055017}
  (\bibinfo{year}{2008}), \eprint{0807.3055}.

\bibitem[{\citenamefont{Aad et~al.}(2011)}]{Aad:2011qa}
\bibinfo{author}{\bibfnamefont{G.}~\bibnamefont{Aad}} \bibnamefont{et~al.}
  (\bibinfo{collaboration}{Atlas}), {``}\bibinfo{title}{{Search for new
  phenomena in final states with large jet multiplicities and missing
  transverse momentum using sqrt(s)=7 TeV pp collisions with the ATLAS
  detector}},{''} (\bibinfo{year}{2011}), \eprint{1110.2299}.

\bibitem[{PAS(2009)}]{PAS-SUS-09-001}
{``}\bibinfo{title}{{Search strategy for exclusive multi-jet events from
  supersymmetry at CMS}},{''} (\bibinfo{year}{2009}), \bibinfo{note}{{CMS PAS
  SUS-09-001}}, \urlprefix\url{http://cdsweb.cern.ch/record/1194509}.

\bibitem[{\citenamefont{Nanopoulos}(2002)}]{Nanopoulos:2002qk}
\bibinfo{author}{\bibfnamefont{D.~V.} \bibnamefont{Nanopoulos}},
  {``}\bibinfo{title}{{F-enomenology}},{''} (\bibinfo{year}{2002}),
  \eprint{hep-ph/0211128}.

\bibitem[{\citenamefont{Harnik et~al.}(2005)\citenamefont{Harnik, Larson,
  Murayama, and Thormeier}}]{Harnik:2004yp}
\bibinfo{author}{\bibfnamefont{R.}~\bibnamefont{Harnik}},
  \bibinfo{author}{\bibfnamefont{D.~T.} \bibnamefont{Larson}},
  \bibinfo{author}{\bibfnamefont{H.}~\bibnamefont{Murayama}}, \bibnamefont{and}
  \bibinfo{author}{\bibfnamefont{M.}~\bibnamefont{Thormeier}},
  {``}\bibinfo{title}{{Probing the Planck scale with proton decay}},{''}
  \bibinfo{journal}{Nucl.Phys.} \textbf{\bibinfo{volume}{B706}},
  \bibinfo{pages}{372} (\bibinfo{year}{2005}), \eprint{hep-ph/0404260}.

\bibitem[{\citenamefont{Giudice and Masiero}(1988)}]{Giudice:1988yz}
\bibinfo{author}{\bibfnamefont{G.}~\bibnamefont{Giudice}} \bibnamefont{and}
  \bibinfo{author}{\bibfnamefont{A.}~\bibnamefont{Masiero}},
  {``}\bibinfo{title}{{A Natural Solution to the mu Problem in Supergravity
  Theories}},{''} \bibinfo{journal}{Phys. Lett.}
  \textbf{\bibinfo{volume}{B206}}, \bibinfo{pages}{480} (\bibinfo{year}{1988}).

\bibitem[{\citenamefont{Babu et~al.}(2003)\citenamefont{Babu, Gogoladze, and
  Wang}}]{Babu:2002tx}
\bibinfo{author}{\bibfnamefont{K.}~\bibnamefont{Babu}},
  \bibinfo{author}{\bibfnamefont{I.}~\bibnamefont{Gogoladze}},
  \bibnamefont{and} \bibinfo{author}{\bibfnamefont{K.}~\bibnamefont{Wang}},
  {``}\bibinfo{title}{{Natural R parity, mu term, and fermion mass hierarchy
  from discrete gauge symmetries}},{''} \bibinfo{journal}{Nucl.Phys.}
  \textbf{\bibinfo{volume}{B660}}, \bibinfo{pages}{322} (\bibinfo{year}{2003}),
  \eprint{hep-ph/0212245}.

\bibitem[{\citenamefont{Ellis et~al.}(2002)\citenamefont{Ellis, Nanopoulos, and
  Olive}}]{Ellis:2001kg}
\bibinfo{author}{\bibfnamefont{J.~R.} \bibnamefont{Ellis}},
  \bibinfo{author}{\bibfnamefont{D.~V.} \bibnamefont{Nanopoulos}},
  \bibnamefont{and} \bibinfo{author}{\bibfnamefont{K.~A.} \bibnamefont{Olive}},
  {``}\bibinfo{title}{{Lower limits on soft supersymmetry breaking scalar
  masses}},{''} \bibinfo{journal}{Phys.\ Lett.} \textbf{\bibinfo{volume}{{\bf
  B525}}}, \bibinfo{pages}{308} (\bibinfo{year}{2002}), \eprint{arXiv:0109288}.

\bibitem[{\citenamefont{Ellis et~al.}(2010)\citenamefont{Ellis, Mustafayev, and
  Olive}}]{Ellis:2010jb}
\bibinfo{author}{\bibfnamefont{J.}~\bibnamefont{Ellis}},
  \bibinfo{author}{\bibfnamefont{A.}~\bibnamefont{Mustafayev}},
  \bibnamefont{and} \bibinfo{author}{\bibfnamefont{K.~A.} \bibnamefont{Olive}},
  {``}\bibinfo{title}{{Resurrecting No-Scale Supergravity Phenomenology}},{''}
  \bibinfo{journal}{Eur. Phys. J.} \textbf{\bibinfo{volume}{C69}},
  \bibinfo{pages}{219} (\bibinfo{year}{2010}), \eprint{1004.5399}.

\bibitem[{\citenamefont{Lopez et~al.}(1993)\citenamefont{Lopez, Nanopoulos, and
  Yuan}}]{Lopez:1992kg}
\bibinfo{author}{\bibfnamefont{J.~L.} \bibnamefont{Lopez}},
  \bibinfo{author}{\bibfnamefont{D.~V.} \bibnamefont{Nanopoulos}},
  \bibnamefont{and} \bibinfo{author}{\bibfnamefont{K.-j.} \bibnamefont{Yuan}},
  {``}\bibinfo{title}{{The Search for a realistic flipped $SU(5)$ string
  model}},{''} \bibinfo{journal}{Nucl. Phys.} \textbf{\bibinfo{volume}{B399}},
  \bibinfo{pages}{654} (\bibinfo{year}{1993}), \eprint{hep-th/9203025}.

\bibitem[{\citenamefont{Barate et~al.}(2003)}]{Barate:2003sz}
\bibinfo{author}{\bibfnamefont{R.}~\bibnamefont{Barate}} \bibnamefont{et~al.}
  (\bibinfo{collaboration}{LEP Working Group for Higgs boson searches}),
  {``}\bibinfo{title}{{Search for the standard model Higgs boson at LEP}},{''}
  \bibinfo{journal}{Phys. Lett.} \textbf{\bibinfo{volume}{B565}},
  \bibinfo{pages}{61} (\bibinfo{year}{2003}), \eprint{hep-ex/0306033}.

\bibitem[{\citenamefont{Yao et~al.}(2006)}]{Yao:2006px}
\bibinfo{author}{\bibfnamefont{W.~M.} \bibnamefont{Yao}} \bibnamefont{et~al.}
  (\bibinfo{collaboration}{Particle Data Group}), {``}\bibinfo{title}{{Review
  of Particle physics}},{''} \bibinfo{journal}{J. Phys.}
  \textbf{\bibinfo{volume}{G33}}, \bibinfo{pages}{1} (\bibinfo{year}{2006}).

\bibitem[{\citenamefont{Komatsu et~al.}(2010)}]{Komatsu:2010fb}
\bibinfo{author}{\bibfnamefont{E.}~\bibnamefont{Komatsu}} \bibnamefont{et~al.}
  (\bibinfo{collaboration}{WMAP}), {``}\bibinfo{title}{{Seven-Year Wilkinson
  Microwave Anisotropy Probe (WMAP) Observations: Cosmological
  Interpretation}},{''} \bibinfo{journal}{Astrophys.J.Suppl.}
  \textbf{\bibinfo{volume}{192}}, \bibinfo{pages}{18} (\bibinfo{year}{2010}),
  \eprint{1001.4538}.

\bibitem[{\citenamefont{Feynman}(1964)}]{Feynman}
\bibinfo{author}{\bibfnamefont{R.~P.} \bibnamefont{Feynman}},
  {``}\bibinfo{title}{{The Feynman Lectures on Physics}},{''}
  (\bibinfo{year}{1964}), \eprint{Volume III}.

\bibitem[{\citenamefont{Martin}(2010)}]{Martin:2009bg}
\bibinfo{author}{\bibfnamefont{S.~P.} \bibnamefont{Martin}},
  {``}\bibinfo{title}{{Extra vector-like matter and the lightest Higgs scalar
  boson mass in low-energy supersymmetry}},{''} \bibinfo{journal}{Phys.Rev.}
  \textbf{\bibinfo{volume}{D81}}, \bibinfo{pages}{035004}
  (\bibinfo{year}{2010}), \eprint{0910.2732}.

\bibitem[{\citenamefont{Huo et~al.}(2011)\citenamefont{Huo, Li, Nanopoulos, and
  Tong}}]{Huo:2011zt}
\bibinfo{author}{\bibfnamefont{Y.}~\bibnamefont{Huo}},
  \bibinfo{author}{\bibfnamefont{T.}~\bibnamefont{Li}},
  \bibinfo{author}{\bibfnamefont{D.~V.} \bibnamefont{Nanopoulos}},
  \bibnamefont{and} \bibinfo{author}{\bibfnamefont{C.}~\bibnamefont{Tong}},
  {``}\bibinfo{title}{{The Lightest CP-Even Higgs Boson Mass in the Testable
  Flipped $SU(5) \times U(1)_X$ Models from F-Theory}},{''}
  (\bibinfo{year}{2011}), \eprint{1109.2329}.

\bibitem[{PAS(2011)}]{PAS-SUS-11-003}
{``}\bibinfo{title}{{Search for supersymmetry in all-hadronic events with
  $\alpha_{\rm T}$}},{''} (\bibinfo{year}{2011}), \bibinfo{note}{{CMS PAS
  SUS-11-003}}, \urlprefix\url{http://cdsweb.cern.ch/record/1370596}.

\bibitem[{\citenamefont{Stelzer and Long}(1994)}]{Stelzer:1994ta}
\bibinfo{author}{\bibfnamefont{T.}~\bibnamefont{Stelzer}} \bibnamefont{and}
  \bibinfo{author}{\bibfnamefont{W.~F.} \bibnamefont{Long}},
  {``}\bibinfo{title}{{Automatic generation of tree level helicity
  amplitudes}},{''} \bibinfo{journal}{Comput. Phys. Commun.}
  \textbf{\bibinfo{volume}{81}}, \bibinfo{pages}{357} (\bibinfo{year}{1994}),
  \eprint{hep-ph/9401258}.

\bibitem[{\citenamefont{Alwall et~al.}(2011)}]{MGME}
\bibinfo{author}{\bibfnamefont{J.}~\bibnamefont{Alwall}} \bibnamefont{et~al.},
  {``}\bibinfo{title}{MadGraph/MadEvent Collider Event Simulation Suite},{''}
  (\bibinfo{year}{2011}), \urlprefix\url{http://madgraph.hep.uiuc.edu/}.

\bibitem[{\citenamefont{Alwall et~al.}(2007)}]{Alwall:2007st}
\bibinfo{author}{\bibfnamefont{J.}~\bibnamefont{Alwall}} \bibnamefont{et~al.},
  {``}\bibinfo{title}{{MadGraph/MadEvent v4: The New Web Generation}},{''}
  \bibinfo{journal}{JHEP} \textbf{\bibinfo{volume}{09}}, \bibinfo{pages}{028}
  (\bibinfo{year}{2007}), \eprint{0706.2334}.

\bibitem[{\citenamefont{Sjostrand et~al.}(2006)\citenamefont{Sjostrand, Mrenna,
  and Skands}}]{Sjostrand:2006za}
\bibinfo{author}{\bibfnamefont{T.}~\bibnamefont{Sjostrand}},
  \bibinfo{author}{\bibfnamefont{S.}~\bibnamefont{Mrenna}}, \bibnamefont{and}
  \bibinfo{author}{\bibfnamefont{P.~Z.} \bibnamefont{Skands}},
  {``}\bibinfo{title}{{PYTHIA 6.4 Physics and Manual}},{''}
  \bibinfo{journal}{JHEP} \textbf{\bibinfo{volume}{05}}, \bibinfo{pages}{026}
  (\bibinfo{year}{2006}), \eprint{hep-ph/0603175}.

\bibitem[{\citenamefont{Conway et~al.}(2009)}]{PGS4}
\bibinfo{author}{\bibfnamefont{J.}~\bibnamefont{Conway}} \bibnamefont{et~al.},
  {``}\bibinfo{title}{PGS4: Pretty Good (Detector) Simulation},{''}
  (\bibinfo{year}{2009}),
  \urlprefix\url{http://www.physics.ucdavis.edu/~conway/research/}.

\bibitem[{\citenamefont{Li et~al.}(2011{\natexlab{m}})\citenamefont{Li, Maxin,
  Nanopoulos, and Walker}}]{cutlhco}
\bibinfo{author}{\bibfnamefont{T.}~\bibnamefont{Li}},
  \bibinfo{author}{\bibfnamefont{J.~A.} \bibnamefont{Maxin}},
  \bibinfo{author}{\bibfnamefont{D.~V.} \bibnamefont{Nanopoulos}},
  \bibnamefont{and} \bibinfo{author}{\bibfnamefont{J.~W.}
  \bibnamefont{Walker}}, {``}\bibinfo{title}{CutLHCO: A Tool For Detector
  Selection Cuts},{''} (\bibinfo{year}{2011}{\natexlab{m}}),
  \urlprefix\url{http://www.joelwalker.net/code/cut_lhco.tar.gz}.

\end{thebibliography}

\end{document}